\PassOptionsToPackage{dvipsnames}{xcolor}
\documentclass{lmcs} 
\pdfoutput=1

\usepackage{lastpage}
\lmcsdoi{18}{1}{30}
\lmcsheading{}{\pageref{LastPage}}{}{}%
{Sep.~29,~2020}{Feb.~04,~2022}{}


\usepackage[utf8]{inputenc}
\usepackage{myisabelle}
\usepackage{subfig}
\usepackage{proof}
\usepackage{hyphenat,xspace,booktabs}
\usepackage{amssymb,amsmath}
\usepackage{prettyref}
\newrefformat{fig}{\textsc{Figure}~\ref{#1}}
\usepackage{tikz}
\usetikzlibrary{backgrounds}
\usetikzlibrary{positioning}
\usetikzlibrary{shapes}
\usetikzlibrary{arrows.meta}
\tikzset{>={Stealth[inset=5.5pt,length=8pt,angle'=45,round]}}
\tikzstyle{every node}=[draw,ellipse]

\newcommand\cLE{\ensuremath\preceq}
\newcommand\SLE{\ensuremath\sqsubseteq}
\newcommand\SGE{\ensuremath\sqsupseteq}
\newcommand\SLT{\ensuremath\sqsubset}

\newcommand\ALL{\isamath{\forall}}%
\newcommand\all{\isamath\bigwedge}%
\newcommand\AND{\isamath\wedge}%
\newcommand\eq{\isamath\equiv}%
\newcommand\EX{\isamath\exists}%
\newcommand\IFF{\isamath\longleftrightarrow}%
\newcommand\IMP{\isamath\Longrightarrow}%
\newcommand\imp{\isamath\longrightarrow}%
\newcommand\IN{\isamath\in}%
\newcommand\OR{\isamath\vee}%
\newcommand\To{\isamath\Rightarrow}%

\newcommand\isalemma{Lemma~\isa}

\newcommand\Nat{\mathbb{N}}
\newcommand\im{\;\grave{~}\,}
\newcommand\defiff{\mathrel{\;\equiv\;}}
\newcommand\defeq{\mathrel{\,\equiv\,}}
\newcommand\CC{\mathcal{C}}
\renewcommand\AA{\mathcal{A}}
\newcommand\XX{\mathcal{X}}
\newcommand\SUP{\bigsqcup}
\newcommand\fq{F}

\makeatletter

\def\tp@#1#2{\@ifnextchar[{\tp@@{#1}{#2}}{\tp@@@{#1}{#2}}}
\def\tp@@#1#2[#3]#4{#3#1\def\mid{\mathrel{#3|}}#4#3#2}
\def\tp@@@#1#2#3{\bgroup\left#1\def\mid{\;\middle|\;}#3\right#2\egroup}
\def\pa{\tp@()}
\def\tp{\tp@()}
\def\set{\tp@\{\}}

\makeatother

\begin{document}

\title{Fixed-Point Theorems for Non-Transitive Relations}
\titlecomment{2012 ACM CCS: Theory of computation $\to$ Logic $\to$ Automated reasoning; 
Theory of computation $\to$ Semantics and reasoning $\to$ Program semantics $\to$ Denotational 
semantics}

\author[J.~Dubut]{J\'er\'emy Dubut}	
\address{National Institute of Informatics, Tokyo, Japan\newline 
Japanese-French Laboratory for Informatics, IRL 3527, Tokyo, Japan}	
\email{\texttt{dubut@nii.ac.jp}}  

\author[A.~Yamada]{Akihisa Yamada}	
\address{National Institute of Advanced Industrial Science and Technology, Tsukuba, Japan
\newline
National Institute of Informatics, Tokyo, Japan}	
\email{\texttt{akihisa.yamada@aist.go.jp}}  






\begin{abstract}
  In this paper,
we develop an Isabelle/HOL library
of order-theoretic fixed-point theorems.
We keep our formalization as general as possible:
we reprove several well-known results about complete orders,
often with only antisymmetry or attractivity, a mild condition implied by either 
antisymmetry or transitivity.
In particular, we generalize various theorems ensuring
the existence of a quasi-fixed point of monotone maps over complete relations,
and show that the set of (quasi-)fixed points is itself complete.
This result generalizes and strengthens theorems of Knaster--Tarski, Bourbaki--Witt, Kleene, Markowsky, Pataraia, Mashburn, Bhatta--George, and Stouti--Maaden.
\end{abstract}

\maketitle

\section*{Introduction}

\emph{Fixed-point theorems} are of fundamental importance in computer science, such as 
in denotational semantics~\cite{scott71} and
in abstract interpretation~\cite{cousot77}, as they allow the definition 
of semantics of loops and recursive functions.
The Knaster--Tarski theorem~\cite{tarski55}
shows that
any monotone map $f : A \to A$ over a complete lattice $(A,\SLE)$ has a fixed point,
and the set of fixed points also forms a complete lattice.
The result was extended in various ways.
\begin{itemize}
\item\emph{Relaxing completeness assumptions:}
Abian and Brown~\cite[Theorem~2]{Abian59}
proved the existence of fixed points under a more general completeness assumption,
which is nowadays called a \emph{weak chain-complete} poset~\cite{BG11}.
Markowsky~\cite{markowsky76}
showed that, for \emph{chain-complete} posets, the set of fixed points are again chain-complete.
Markowsky's proof uses the Bourbaki--Witt theorem (see below), whose original proof is non-elementary in the sense that it relies on ordinals and Hartogs' theorem. 
Pataraia~\cite{pataraia97} 
gave an elementary proof of the existence of least fixed points for
\emph{pointed directed-complete} posets.

\item\emph{Relaxing order assumptions:}
Fixed points are studied also for \emph{pseudo-orders}~\cite{trellis},
relaxing transitivity.
Bhatta and George~\cite{Bhatta05,BG11} gave a non-elementary proof showing that
the set of fixed points over weak chain-complete pseudo-orders
is again weak chain-complete.
Stouti and Maaden~\cite{SM13} showed that every monotone map over a complete pseudo-order has a (least) fixed point,
with an elementary proof.

\item\emph{Alternative to monotonicity:}
Another line of research on fixed points is to consider \emph{inflationary} maps rather than monotone ones.
The Bourbaki--Witt theorem~\cite{bourbaki49}
states that any inflationary map over a chain-complete poset has a fixed point, and its proof is non-elementary as already mentioned.
Abian and Brown~\cite[Theorem~3]{Abian59} also gave an elementary proof for a generalization of the Bourbaki--Witt theorem applied to weak chain-complete posets.

\item\emph{Iterative approach:}
One last line of research on fixed points we would like to mention is the \emph{iterative} approach.
Kantorovitch showed that
for any \emph{$\omega$-continuous} map $f$
over a complete lattice,\footnote{
More precisely,
he assumes a conditionally complete lattice defined over vectors and that
$\bot \SLE f\:\bot$ and $f\:v' \SLE v'$.
Hence $f$, which is monotone,
is a map over the complete lattice $\{v \mid \bot \SLE v \SLE v'\}$.
}
the
iteration $\bot, f\:\bot,f^2\:\bot,\dots$ converges to a fixed point
\cite[Theorem~I]{kantorovitch39}.
Tarski~\cite{tarski55} also claimed a similar result for a
\emph{countably distributive} map over a
\emph{countably complete} lattice.
Kleene's fixed-point theorem states that,
for \emph{Scott-continuous} maps over pointed directed-complete posets,
the iteration converges to the least fixed point.
Finally, Mashburn~\cite{mashburn83} proved a version for 
$\omega$-continuous maps over $\omega$-complete posets,
which covers Kantorovitch's, Tarski's and Kleene's results.

\end{itemize}

In this paper,
we formalize these fixed-point theorems in a general form,
using the proof assistant \emph{Isabelle/HOL}~\cite{Isabelle}.
The use of proof assistants
such as Coq~\cite{coq}, Agda~\cite{agda}, HOL-Light~\cite{hol-light}, and Isabelle/HOL,
are exemplified prominently by
a proof of the four-colour theorem in Coq~\cite{4color},
a proof of the Kepler conjecture in discrete geometry in HOL-Light and Isabelle~\cite{flyspeck}, a formal verification of an OS microkernel in Isabelle/HOL~\cite{sel4}, etc.,
where proofs are so big that human reviewing would not be able to verify the correctness of the proofs within a reasonable time.
In this work, we utilize another aspect of proof assistants:
they are also engineering tools for developing mathematical theories.
In particular, Isabelle/JEdit~\cite{isabelle/jedit} is
a \emph{very} smart environment for developing theories in Isabelle/HOL.
There, the proofs we write are checked ``on the fly'',
so that one can easily refine proofs or even theorem statements by just changing a part of it and see if Isabelle complains or not.
Sledgehammer~\cite{sledgehammer} can often automatically fill relatively small gaps in proofs
so that we can concentrate on more important aspects.
Isabelle's counterexample finders~\cite{quickcheck,nitpick}
should also be highly appreciated,
considering the amount of time one would spend trying in vain to prove a false claim.

We adopt an \emph{as-general-as-possible} approach:
all theorems
are proved without assuming the underlying relations to be orders.
One can easily find several formalizations of complete partial orders or lattices in Isabelle's standard library.
They are, however, defined on partial orders
and thus not directly reusable for general relations.

In particular, we provide the following:
\begin{itemize}
\item Several \emph{locales}~\cite{Kammuller00,locale} that help organizing the different order-theoretic conditions, 
such as reflexivity, transitivity, antisymmetry, and their combinations, as well as concepts such as connex and well-related sets, analogues of chains 
and well-ordered sets in a non-ordered context
(\prettyref{sec:prelim}).
\item Existence of fixed points:
We provide two proof methods for proving that a monotone or inflationary mapping 
$f : A \to A$ over a complete related set $\tp{A,\SLE}$
has a \emph{quasi-fixed point} $f\:x \sim x$,
meaning $x \SLE f\:x \mathrel\land f\:x \SLE x$, for various notions of completeness.
The first one (\prettyref{sec:knaster-tarski}), similar to the proof by Stouti and Maaden~\cite{SM13}, 
does not require any ordering assumptions,
but relies on completeness with respect to all subsets.
The second one (\prettyref{sec:weak_chain}),
inspired by a \emph{constructive} approach by Grall~\cite{grall10},
is a proof method based on the notion of derivations.
For this method, we demand antisymmetry 
(to avoid the necessity of the axiom of choice),
and the statement can then be instantiated to \emph{well-complete} sets,
a generalization of weak chain-completeness.
This also allows us to generalize the Bourbaki--Witt theorem~\cite{bourbaki49} to pseudo-orders.
\item Completeness of the set of fixed points (\prettyref{sec:completeness}): We further show that
if $(A,\SLE)$ satisfies a mild condition, which we call \emph{attractivity} and
which is implied by either transitivity or antisymmetry,
then the set of quasi-fixed points inherits the completeness class of
$(A,\SLE)$, if it is at least well-complete.
The result instantiates to
the full completeness (generalizing Knaster--Tarski and~\cite{SM13}), 
directed-completeness~\cite{pataraia97},
chain-completeness~\cite{markowsky76}, 
and weak chain-completeness~\cite{BG11}.
\item Iterative construction (\prettyref{sec:kleene}):
For an $\omega$-continuous map over an $\omega$-complete related set,
we show that suprema of $\set{f^n\:\bot \mid n\in\Nat}$ are quasi-fixed points.
Under attractivity, the quasi-fixed points obtained from this method 
are precisely the least quasi-fixed points of $f$.
This generalizes Mashburn's result, and thus ones by
Kantorovitch, Tarski and Kleene.
\end{itemize}

The formalization
is available in the \emph{Archive of Formal Proofs}~\cite{AFP}.
We can easily ensure that our development indeed does not use the axiom of choice,
by the fact that Isabelle validates the proofs only by loading basic HOL libraries, excluding the axiom of choice (\isa{HOL.Hilbert_Choice}).

We remark that all these results would have required much more effort than we spent
(if possible at all),
if we were not with the aforementioned smart assistance of Isabelle.
Our workflow was often the following: first we formalize existing proofs, try relaxing assumptions, 
see where the proof breaks, and at some point ask for a counterexample.
We also observe that a carefully chosen use of notations and locales lets us do mathematics in Isabelle without going too far beyond daily mathematics.

\paragraph*{Comparison with~\cite{yamada19}}~
The present paper is built upon authors' work~\cite{yamada19} presented at ITP'19,
but the entire formalization went through an overhaul.
Formalizations of \prettyref{sec:wos}, the proof of existence of quasi-fixed points using well-completeness (\prettyref{sec:weak_chain}), and most of the proof of completeness of the set of 
(quasi-)fixed points (\prettyref{sec:completeness}) are new materials. The rest has been 
accommodated to fit with this new material, as well as to make better notations, proof 
structures, etc.

\section{Preliminaries}
\label{sec:prelim}

We develop our theory in Isabelle/HOL and present statements following its notation.
Here we briefly explain notions and notations needed for the paper.
We refer interested readers to the textbook~\cite{Isabelle} for more detail. 
In Isabelle,
\isa{\IMP} and \isa{\imp} denote the logical implication.\footnote{
Technical difference between their behaviors can be ignored for reading the paper.
}
Function application is written $f\:x$.
By \isa{$A$ :: 'a set} we denote a set $A$ whose elements are of type \isa{'a},
and \isa{$R$ :: 'a \To 'a \To bool} is a binary predicate defined over \isa{'a}.
Type annotations ``\isa{::~_}'' are omitted unless necessary.

Now we introduce several notions that will be 
needed to state and prove fixed-point theorems.
We call the pair $\tp{A,\SLE}$ of a set $A$ and a binary relation $(\SLE)$ over $A$ a \emph{related set}.
One could also call it a \emph{graph} or an \emph{abstract reduction system},
but then some terminologies like ``complete'' become incompatible.
A map $f : I \to A$ over related sets from $\tp{I,\cLE}$ to $\tp{A,\SLE}$
is \emph{relation preserving}, or \emph{monotone},
if $i \cLE j$ implies $f\:i \SLE f\:j$.
We define this property, in particular restricted to the set $I$,
in Isabelle as follows:
\begin{isabelle}
\definition "monotone_on $I$ (\cLE) (\SLE) $f$ \eq \ALL{$i$} \IN $I$. \ALL{$j$} \IN $I$. $i$ \cLE\ $j$ \imp\ $f$ $i$ \SLE\ $f$ $j$"
\end{isabelle}
Hereafter, in our Isabelle code, we use symbols
\isa{(\SLE)} denoting a variable of type \isa{'a \To 'a \To bool},
and \isa{(\cLE)} denoting a variable of type \isa{'b \To 'b \To bool}.
More precisely, statements and definitions using these symbols are made in a \emph{context} which fixes a binary relation and introduces an infix notation for it:
\begin{isabelle}
\context \fixes less_eq :: "'a \To 'a \To bool" (\infix "\SLE" 50)
\end{isabelle}
For clarity, we explicitly write the relations \isa{(\cLE)} or \isa{(\SLE)} as 
parameters in the definitions.

Other core ingredients in fixed-point theorems are
the \emph{least upper bounds (suprema)} and \emph{greatest lower bounds (infima)}.
The predicates for being upper/lower bounds and greatest/least elements
are defined as follows:
\begin{isabelle}
\definition "bound $X$ (\SLE) $b$ \eq \ALL{$x$} \IN $X$. $x$ \SLE $b$"
\definition "extreme $X$ (\SLE) $e$ \eq $e$ \IN $X$ \AND (\ALL{$x$} \IN $X$. $x$ \SLE $e$)"
\end{isabelle}
Note that we chose such constant names that do not suggest which side is greater or lower.
Thus the suprema and infima are uniformly defined as follows:
\begin{isabelle}
\abbreviation "extreme_bound $A$ (\SLE) $X$ \eq extreme \{$b$ \IN $A$. bound $X$ (\SLE) $b$\} (\SGE)"
\end{isabelle}
Hereafter, we write \isa{(\SGE)} for the dual of $(\SLE)$:
$x \SGE y \equiv y \SLE x$,
and \isa{\{$x$ \IN $A$. $P$ $x$\}} is one of the Isabelle/HOL notations for set comprehension, $\{x \in A \mid P\ x\}$ in daily mathematics.

We can already prove some useful lemmas. For instance,
if $f : I \to A$ is relation preserving and
$I$ has a greatest element $e \in I$,
then $f\:e$ is a supremum of the image of $I$ by $f$, denoted by $f \im I$ following Isabelle 
notations.
Note here that no assumption is imposed on the relations $(\cLE)$ and $(\SLE)$.
\begin{isabelle}
\lemma monotone_extreme_imp_extreme_bound:
  \assumes "$f \im I \subseteq A$" \and "monotone_on $I$ (\cLE) (\SLE) $f$" \and "extreme $I$ (\cLE) $e$"
  \shows "extreme_bound $A$ (\SLE) ($f \im I$) ($f\:e$)"
\end{isabelle}

\subsection{Locale Hierarchy of Relations}

We now define basic properties of binary relations,
in form of \emph{locales}~\cite{Kammuller00,locale}.
Isabelle's locale mechanism allows us to conveniently manage notations,
assumptions and facts.
For instance, we introduce the following locale for infix notation of a 
related set.

\begin{isabelle}
\locale related_set =
  \fixes $A$ :: "'a set" \and less_eq :: "'a \To 'a \To bool" (\infix "\SLE" 50)
\end{isabelle}

The most important feature of locales is that we can impose assumptions on parameters.
For instance, we define a locale for reflexive relations as follows.
\begin{isabelle}
\locale reflexive = related_set +
  \assumes refl[intro]: "$x$ \IN $A$ \IMP $x$ \SLE $x$"
\end{isabelle}
This declaration is logically equivalent to defining predicate \isa{"reflexive"}
with the following equation:
\begin{isabelle}
  reflexive_def: "reflexive $A$ (\SLE) \eq \ALL{$x$}. $x$ \IN $A$ \imp $x$ \SLE $x$"
\end{isabelle}
Compared to just defining a predicate,
declaring a locale will introduce a named context where we can collect facts and give them attributes to guide Isabelle's automation when proving theorems in the locale.
For instance, the ``\isa{[intro]}'' attribute above instructs Isabelle to use the assumption \isa{refl} as an introduction rule in proof automation.
Below are some examples proved in locale \isa{reflexive}:
\begin{isabelle}
\lemma (\iin reflexive) extreme_singleton[simp]: "$x$ \IN $A$ \IMP extreme \{$x$\} (\SLE) $y$ \IFF $x$ = $y$"
\end{isabelle}
\begin{isabelle}
\lemma (\iin reflexive) extreme_bound_singleton: "$x$ \IN $A$ \IMP extreme_bound $A$ (\SLE) \{$x$\} $x$"
\end{isabelle}

Similarly we define transitivity and antisymmetry:
\begin{isabelle}
\locale transitive = related_set +
  \assumes trans[trans]: "$x$ \SLE $y$ \IMP $y$ \SLE $z$ \IMP $x$ \IN $A$ \IMP $y$ \IN $A$ \IMP $z$ \IN $A$ \IMP $x$ \SLE $z$"
\end{isabelle}
\begin{isabelle}
\locale antisymmetric = related_set +
  \assumes antisym: "$x$ \SLE $y$ \IMP $y$ \SLE $x$ \IMP $x$ \IN $A$ \IMP $y$ \IN $A$ \IMP $x$ = $y$"
\end{isabelle}

Another merit of using locales is that
it is straightforward to combine assumptions.
Some well-known combinations are
\emph{quasi-ordered} (also sometimes called pre-ordered) sets for reflexive and transitive relations
and \emph{partially ordered} sets (posets) for antisymmetric quasi-ordered sets.
\begin{isabelle}
\locale quasi_ordered_set = reflexive + transitive
\end{isabelle}
\begin{isabelle}
\locale partially_ordered_set = quasi_ordered_set + antisymmetric
\end{isabelle}

A less known but convenient assumption is being a \emph{pseudo-order},
coined by Skala~\cite{trellis} for reflexive and antisymmetric relations.
There, the supremum of a singleton set $\set{x}$ uniquely exists---$x$ itself.
\begin{isabelle}
\locale pseudo_ordered_set = reflexive + antisymmetric
\end{isabelle}

\begin{isabelle}
\lemma (\iin pseudo_ordered_set) extreme_bound_singleton_eq[simp]:
  "$x$ \IN $A$ \IMP extreme_bound $A$ (\SLE) \{$x$\} $y$ \IFF $x$ = $y$" 
\end{isabelle}
It is clear that a partial order is also a pseudo-order,
which is stated by the following \emph{sublocale} declaration.
\begin{isabelle}
\sublocale partially_ordered_set $\subseteq$ pseudo_ordered_set
\end{isabelle}
This declaration is logically equivalent to proving the fact:
\begin{isabelle}
  "partially_ordered_set $A$ (\SLE) \IMP pseudo_ordered_set $A$ (\SLE)"
\end{isabelle}
The difference is that, after the sublocale declaration, facts proved in \isa{pseudo_ordered_set} will be automatically available in \isa{partially_ordered_set}.

Although these combinations are sufficient for the rest of this paper,
we also present all locales combining these basic properties
and their relationships in \prettyref{fig:non-orders}.

\begin{figure}
\small
\def\t{-2}
\def\a{4}
\def\at{2}
\def\s{-4}
\def\st{-6}
\begin{tikzpicture}[scale = 1.1]
\tikzstyle{every edge}=[draw]
\draw
 (0,0) node (rel) {\isa{related_set}}
 (\t,1) node (trans) {\isa{transitive}}
 (0,-2) node (refl) {\isa{reflexive}}
 (0,2) node (irr) {\isa{irreflexive}}
 (\s,0) node (sym) {\isa{symmetric}}
 (\a,0) node (anti) {\isa{antisymmetric}}
 (\at,1) node (near) {\isa{near_order}}
 (\a,2) node (asym) {\isa{asymmetric}}
 (\a,-2) node (pso) {\isa{pseudo_order}}
 (\at,-1) node (po) {\isa{partial_order}}
 (\t,-1) node (qo) {\isa{quasi_order}}
 (0,3) node (str) {\hspace{3em}\isa{strict_order}\mbox{\hspace{3em}}}
 (\st,-1) node (equiv) {\isa{equivalence}}
 (\st,1) node (peq) {\isa{partial_equivalence}}
 (\st,3) node (emp) {$\emptyset$}
 (\s,-2) node (tol) {\isa{tolerance}}
 (\s,2) node (ntol) {$\neg$\isa{tolerance}}
 ;
\draw[->]
 (near) edge[color=blue] ([xshift=56,yshift=-7]str)
 (irr) edge[color=red] ([xshift=-50,yshift=-8]str)
 (trans) edge[color=blue] ([xshift=-56,yshift=-7]str)
 (asym) edge[color=red] ([xshift=60,yshift=-7]str)
 (trans) edge[color=green!70!black] (near)
 (anti) edge[color=red] (near)
 (irr) edge[color=green!70!black] (asym)
 (anti) edge[color=blue] (asym)
 (anti) edge[color=blue] (pso)
 (near) edge[color=blue] (po)
 (pso) edge[color=red] (po)
 (refl) edge[color=green!70!black] (pso)
 (trans) edge[color=blue] (qo)
 (refl) edge[color=red] (qo)
 (qo) edge[color=green!70!black] (po)
 (qo) edge[color=green!70!black] (equiv)
 (peq) edge[color=blue] (equiv)
 (trans) edge[color=green!70!black] (peq)
 (peq) edge[color=blue] (emp)
 (str) edge[color=green!70!black] (emp)
 (sym) edge[color=red] (peq)
 (sym) edge[color=blue] (tol)
 (sym) edge[color=blue] (ntol)
 (irr) edge[color=green!70!black] (ntol)
 (refl) edge[color=green!70!black] (tol)
 (ntol) edge[color=red] (emp)
 (tol) edge[color=red] (equiv);
\draw[line width=2pt,-{Stealth[inset=9pt,length=14pt,angle'=45,round]}]
 (rel) edge[color=red] (trans)
 (rel) edge[color=blue] (irr)
 (rel) edge[color=blue] (refl)
 (rel) edge[color=green!70!black] (sym)
 (rel) edge[color=green!70!black] (anti)
 ;
\end{tikzpicture}
\caption{\label{fig:non-orders}
Combinations of basic properties.
The five outgoing arrows from \isa{related\_sets} indicate atomic assumptions.
We do not present the combination
of \isa{reflexive} and \isa{irreflexive}, which is empty,
and one of \isa{symmetric} and \isa{antisymmetric},
which is a subset of equality.
Node ``$\neg$\isa{tolerance}'' indicates the negated relation is \isa{tolerance},
and ``$\emptyset$'' is the empty relation.
}
\end{figure}

Readers already familiar with Isabelle/HOL might question why we use locales instead of \emph{classes}.
Indeed, Isabelle/HOL already has a class that introduces the order symbol \isa{$\le$}.
One of the drawbacks of this approach is that we cannot restrict our interest to the set $A$ but we are forced to work with \isa{UNIV}.
Another drawback is that one type must have one order, which forbids our results to be instantiated to other relations on the same type.
Our approach, making the relation of concern explicit as an argument,
is sometimes called the \emph{dictionary-passing} style~\cite{codegen}.
On one hand this design choice adds a notational burden,
but on the other hand
it allows instantiating results to arbitrary relations over a type,
for which the class mechanism fixes one ordering.
In the formalization we also import our results into the class hierarchy, by taking \isa{$A$ = UNIV} and \isa{(\SLE) = ($\le$)}.

\subsection{Well Related Sets}
\label{sec:wos}

A \emph{well-ordered set} is a poset $\tp{A,\SLE}$ such that
every nonempty subset of $A$ has a least element.
We generalize the notion to \emph{well-related set},
which does not assume posets:
\begin{isabelle}
\locale well_related_set = related_set +
  \assumes "$X$ $\subseteq$ $A$ \IMP $X$ $\neq$ \{\} \IMP \EX{$e$}. extreme $X$ (\SGE) $e$"
\end{isabelle}
Every well-related set is \emph{connex},
i.e., any two elements are comparable.
\begin{isabelle}
\locale connex = related_set +
  \assumes "$x$ \IN $A$ \IMP $y$ \IN $A$ \IMP $x$ \SLE $y$ \OR $y$ \SLE $x$"
\end{isabelle}

\begin{isabelle}
\sublocale well_related_set $\subseteq$ connex
\end{isabelle}
\begin{proof}
Let $x,y \in A$.
The set $\set{x,y}$ has a least element,
so $x \SLE y$ or $y \SLE x$.
\end{proof}

It is also easy to see that connexity implies reflexivity:
\begin{isabelle}
\sublocale connex $\subseteq$ reflexive
\end{isabelle}

A crucial observation is that every well-related set is \emph{well-founded},
that is,
the \emph{asymmetric part} of $(\SLE)$ defined by
$x \SLT y \defiff x \SLE y \land y \not\SLE x$
satisfies the induction principle:
\begin{isabelle}
  "$\forall a \in A$. ($\forall x \in A$. ($\forall y \in A$. $y$ \SLT $x$ \imp $P$ $y$) \imp $P$ $x$) \imp $P$ $a$"
\end{isabelle}
The proof is easy, using the classical result
that well-foundedness is equivalent to assuming
that every nonempty $X \subseteq A$ has a minimal element;
least elements are also minimal.

We remark that under antisymmetry,
well-relatedness and well-orderedness are equivalent.
We just define well-ordered sets as antisymmetric well-related sets,
and prove that they are actually posets.
\begin{isabelle}
\locale well_ordered_set = antisymmetric + well_related_set
\end{isabelle}
\begin{isabelle}
\sublocale well_ordered_set $\subseteq$ partially_ordered_set
\end{isabelle}
\begin{proof}
Since well-related sets are connex and thus reflexive,
and since we explicitly assume antisymmetry,
it only remains to show that $\tp{A,\SLE}$ is transitive.%
\footnote{This elegant proof of transitivity is contributed by an anonymous reviewer.}
So fix $x$, $y$ and $z \in A$ with $x \SLE y$ and $y \SLE z$, and let us 
prove that $x \SLE z$.
By well-relatedness, the set $\set{x,y,z}$ has an extreme element $l$.
There are three possible cases:
\begin{itemize}
	\item If $l = x$, then by extremality $x = l \SLE z$.
	\item If $l = y$, then by extremality $y = l \SLE x$, and by antisymmetry $x = y \SLE z$.
	\item If $l = z$, then by extremality $z = l \SLE y$, and by antisymmetry $x \SLE y = z$.\qedhere
\end{itemize}
\end{proof}

\section{Existence of Fixed Points in Complete Related Set}
\label{sec:knaster-tarski}

A related set $\tp{A,\SLE}$ is $\CC$-\emph{complete},
where $\CC$ is a class of sets,
if every subset $X \subseteq A$ belonging to $\CC$ has a supremum in $A$.
\begin{isabelle}
\definition complete ("_-complete"[999]1000) \where
  "$\CC$-complete $A$ (\SLE) \eq \ALL{$X$} $\subseteq$ $A$. $X$ \IN $\CC$ \imp (\EX{$s$}. extreme_bound $A$ (\SLE) $X$ $s$)"
\end{isabelle}

In this section we focus on the strongest completeness assumption \isa{UNIV-complete},
i.e., any subset of elements has a (not necessarily unique) supremum,
and further generalize 
Stouti and Maaden's result so that it works on complete related sets,
relaxing even reflexivity and antisymmetry.
Much as in the Bourbaki--Witt theorem, we also generalize the 
monotonicity assumption to allow inflationary maps, that is, maps such that
$x \SLE f\:x$ for all $x$.
\label{sec:qfp-exists}

Notice that \isa{UNIV-complete} does not explicitly demand infima,
in Isabelle, \isa{"\EX{$i$}. extreme_bound $A$ (\SGE) $X$ $i$"}.
This is a well-known consequence in complete lattices, namely that infima can be 
defined in terms of suprema as greatest lower bounds, and
luckily the proof does not rely on any property of orders. This allows us to 
state that \isa{UNIV-complete} is auto-dual in the following sense:
\begin{isabelle}
\lemma complete_dual:
  \assumes "UNIV-complete $A$ (\SLE)" \shows "UNIV-complete $A$ (\SGE)"
\end{isabelle}

In the rest of the section, our goal is to prove that a monotone or inflationary map on an \isa{UNIV-complete} set has a fixed point, following closely the 
proof by Stouti and Maaden~\cite{SM13}. The structure will be the same as 
their proof, only accommodating some arguments to fit our general 
framework.

First we just assume completeness
and analyze the existence of fixed points.
Fortunately, Quickcheck~\cite{quickcheck}
quickly refutes the existence of \emph{strict} fixed point $f\:x = x$
even when $f$ is monotone and inflationary.
\begin{exa}[by Quickcheck]
Let $A = \set{a_1,a_2}$, $(\SLE) = A \times A$,
$f\:a_1 = a_2$, and $f\:a_2 = a_1$.
$f$ is monotone and inflationary but $f\:x \neq x$ for either $x \in A$.
\end{exa}
Hence, we instead show the existence of a \emph{quasi-fixed point} $f\:x \sim x$, that is, 
$f\:x \SLE x$ and 
$x \SLE f\:x$.
The set of quasi-fixed points is included in the set of fixed points for 
antisymmetric relations -- the inclusion can be strict without reflexivity;
hence the Stouti--Maaden theorem is further generalized by relaxing reflexivity.
Moreover,
we develop an existence theorem that generalizes both monotone and inflationary $f$,
namely,
quasi-fixed points exist if
$f : A \to A$ is \emph{monotone or inflationary at each point}:
\[
\forall x \in A.\ x \SLE f\:x \mathrel\vee (\forall y \in A.\ y \SLE x \longrightarrow f\:y \SLE f\:x)
\]
We develop proofs within the following locale,%
\footnote{%
The assumption $f\im A \subseteq A$ could be equivalently written \isa{$f$ : $A\to A$} in Isabelle;
unfortunately, the latter notation in the Isabelle/HOL library automatically enables the axiom of choice.
} so that we can refer to them in the proofs of later theorems:
\begin{isabelle}
\locale fixed_point_proof = related_set +
  \fixes $f$ \assumes "$f$ \` $A$ $\subseteq$ $A$"
\end{isabelle}

We follow Stouti and Maaden's proof~\cite{SM13};
one of their insights is in considering the set of subsets of $A$ that are closed under $f$
and themselves ``complete'':
\begin{isabelle}
\definition $\AA$ \where "$\AA$ \eq
  \{$B$. $B \subseteq A$ \AND $f \im B \subseteq B$ \AND (\ALL$X s$. $X \subseteq B$ \imp extreme_bound $A$ (\SLE) $X$ $s$ \imp $s$ \IN $B$)\}"
\end{isabelle}
Here we slightly modified Stouti and Maaden's definition:
by a ``complete'' subset $B \subseteq A$
we mean that \emph{any} supremum with respect to $\tp{A,\SLE}$ is in $B$,
since suprema are not necessarily unique without antisymmetry.
We denote the intersection of all those subsets by $C$:
\begin{isabelle}
\definition $C$ \where "$C$ \eq $\bigcap$ $\AA$"
\end{isabelle}
and show that
a supremum of $C$,
which exists due to completeness, is a quasi-fixed point.
The proof basically follows that by Stouti and Maaden,
but after formalizing their proof we noticed that the monotonicity condition can be generalized with a tiny modification.
\begin{isabelle}
\lemma  qfp_as_extreme_bound:
  \assumes "$\forall x \in A.\ x \SLE f\:x \mathrel\vee (\forall y \in A.\ y \SLE x \longrightarrow f\:y \SLE f\:x)$"
    \and "extreme_bound $A$ (\SLE) $C$ $c$"
  \shows "$f$ $c$ $\sim$ $c$"
\end{isabelle}

\begin{proof}
First, observe that $C \in \AA$. Indeed: 
\begin{itemize}
	\item $C \subseteq A$: since $A$ is closed under $f$, $A \in \AA$.
	\item $f \im C \subseteq C$:
  for every $B \in \AA$,
  we have $f \im C \subseteq f \im B \subseteq B$.
  So $f \im C \subseteq \bigcap \AA = C$.
	\item completeness: given $X \subseteq C$ and its supremum $s$ in $A$,
	we prove $s \in C$, that is,
	$s \in B$ for every $B \in \AA$. 
	Indeed, we have $X \subseteq C \subseteq B$ and the completeness of $B$ ensures $s \in B$. 
\end{itemize}
This implies that $c \in C$. Moreover, since $f \im C \subseteq C$,
we have $f\:c \in C$, and since $c$ is a supremum of $C$, we get $f\:c \SLE c$.
It remains to prove the converse orientation $c \SLE f\:c$. This inequality is obvious when 
$f$ is inflationary at $c$, so let us focus on the case when $f$ is monotone at $c$, that is, $\forall d \in A.\ d \SLE c \imp f\:d \SLE f\:c$.
To this end we consider the following set $D$:
\begin{isabelle}
\define $D$ \where "$D$ \eq \{$x$ \IN $C$. $x$ \SLE $f$ $c$\}"
\end{isabelle}
We conclude by proving that $D \in \AA$, since this implies $C \subseteq D$
and in particular $c \in D$, which means $c \SLE f\:c$.
\begin{itemize}
	\item $D \subseteq A$: because $D \subseteq C \subseteq A$.
	\item $f \im D\subseteq D$: Let $d \in D$. So $d \in C$, and also 
	$f\:d \in C$ since $f \im C \subseteq C$. 
	Furthermore, since $c$ is a supremum of $C$, 
	we have $d \SLE c$. With the monotonicity assumption we get $f\:d \SLE f\:c$ and thus 
	$f\:d \in D$.
	\item completeness: Given $E \subseteq D$ and its supremum $s$ in $A$,
	we prove that $s \in D$. 
	Since $E \subseteq D \subseteq C$, then by completeness of $C$, 
	$s \in C$.
	Additionally, since $E \subseteq D$, $f\:c$ is a bound of $E$,
	and as $s$ is a least of such, $s \SLE f\:c$, that is $s \in D$.
\qedhere
\end{itemize}
\end{proof}

This general lemma allows us to conclude that if $\tp{A,\SLE}$ is complete for a 
notion of completeness that includes the subset $C$, then $f$ has a 
quasi-fixed point given by the existing supremum $c$ of $C$. This is enforced in particular when $\tp{A,\SLE}$ is \isa{UNIV}-complete:
\begin{isabelle}
\theorem complete_infl_mono_imp_ex_qfp:
  \assumes "UNIV-complete $A$ (\SLE)" \and "$\forall x \in A.\ x \SLE f\:x \mathrel\vee (\forall y \in A.\ y \SLE x \longrightarrow f\:y \SLE f\:x)$"
  \shows "$\exists p \in A$. $f\:p \sim p$"
\end{isabelle}
This result generalizes one in our previous work~\cite{yamada19},
where the monotonicity condition is generalized so that inflationary maps are also covered.
It is easy to see that this result indicates the existence of a strict fixed point
if $\tp{A,\SLE}$ is antisymmetric and \isa{UNIV}-complete.
The result covers Stauti and Maaden's existence theorem,
with generalized monotonicity condition and without the reflexivity assumption.
\begin{isabelle}
\corollary (\iin antisymmetric) complete_infl_mono_imp_ex_fp:
  \assumes "UNIV-complete $A$ (\SLE)" \and "$\forall x \in A.\ x \SLE f\:x \mathrel\vee (\forall y \in A.\ x \SLE y \longrightarrow f\:x \SLE f\:y)$"
  \shows "$\exists p \in A$. $f\:p = p$"
\end{isabelle}

\section{Fixed Points in Well-Complete Antisymmetric Sets}
\label{sec:weak_chain}

Let us say that
a related set $(A,\SLE)$ is \emph{well-complete}
if every well-related subset of $A$, including the empty set, has a supremum.
In Isabelle,
\begin{isabelle}
\abbreviation "well_complete $A$ (\SLE) $\equiv$ {\{$X$. well_related_set $X$ (\SLE)\}-complete} $A$ (\SLE)"
\end{isabelle}
Well-completeness is a generalization of
\emph{weak chain-completeness} (named so in~\cite{Bhatta05}, but 
already used in~\cite{Abian59}),
which assumes that every well-ordered subset has a supremum.
Recall that in the presence of antisymmetry,
well-relatedness and well-orderedness coincide, and that so do 
well-completeness and weak chain-completeness.
In this section, we prove that every 
inflationary or monotone map over a well-complete antisymmetric set
has a fixed point. This generalizes Bhatta and George's existence of 
fixed points~\cite{Bhatta05} by removing reflexivity. This result will be 
further generalized in~\prettyref{sec:completeness}.

In order to formalize such a theorem in Isabelle,
we followed Grall's~\cite{grall10} elementary proof for Bourbaki--Witt and Markowsky's theorems.
His idea is to consider well-founded ``derivation trees'' over $A$,
where from a set $C \subseteq A$ of premises
one can ``derive'' $f\:(\SUP C)$ if $C$ is a chain.
The main observation is as follows:
Let $D$ be the set of all the derivable elements; that is,
for each $d \in D$ there exists a well-founded derivation
whose root is $d$.
It is shown that $D$ is a chain,
and hence one can build a derivation yielding $f\:(\SUP D)$,
and $f\:(\SUP D)$ is shown to be a fixed point. This idea is also very 
similar to the proof in~\cite{Abian59}, where the notion of $a$-chain is 
analogue to derivations in Grall's proof.

\begin{figure}
\hfil
\subfloat[A well-founded derivation\label{fig:wfd}]{\hspace{20pt}
$\infer{f^\omega\bot}{
  \bot&
  \infer{f^2\bot}{\infer{f\bot}{\bot}}&
  \infer*{f^4\bot}{\bot}&
  \infer*{f^6\bot}{\bot}&
  \dots
}$\hspace{20pt}
}\hfil
\subfloat[The unique well-order derivation\label{fig:wod}]{
\hspace{20pt}
$\bot \SLT f\bot \SLT f^2\bot \SLT \dots \SLT f^\omega\bot$
\hspace{20pt}}\hfil
\caption{Approaches for deriving
 $f^\omega\bot = \SUP\{f^i\bot \mid i\in\Nat\}$
}
\end{figure}
We started formalizing his proof smoothly in Isabelle/HOL,
until the point of building a derivation tree containing all derivable elements.
There, it appears to us that the axiom of choice is necessary:
we need to choose one derivation for each derivable element,
and then aggregate into one derivation.
Note that a derivable element may have infinitely many well-founded derivations (\prettyref{fig:wfd}).

Of course, the axiom of choice is available in Isabelle/HOL,
but we found a way to avoid using it.
We utilize the following lemma, stating that the union of (infinitely many)
downward-closed well-founded sets
is well-founded.
\begin{isabelle}
\lemma closed_UN_well_founded:
  \assumes "\ALL{$X$} \IN $\XX$. well_founded $X$ (\SLT) \AND (\ALL{$x$} \IN $X$. \ALL{$y$} \IN $\bigcup\XX$. $y \SLT x \imp y \in X$)"
  \shows "well_founded ($\bigcup\XX$) (\SLT)"
\end{isabelle}
\begin{proof}
We show that any nonempty $S \subseteq \bigcup \XX$ has a minimal element.
Let $x \in S$. Then there exists $X \in \XX$ such that $x \in X$.
Due to the assumption on $\XX$, $(X,\SLT)$ is well-founded.
Hence, since $S \cap X \subseteq X$ is nonempty containing $x$,
$S \cap X$ has a minimal element $z$.
We show that $z$ is also minimal in $S$ by contradiction.
So suppose that $y \in S$ with $y \SLT z$ exists.
Since $y \in S \subseteq \bigcup \XX$,
by the assumption on $\XX$ and $z \in X$ we get $y \in X$.
Then with $y \in S$ we get $y \in S \cap X$ and $y \SLT z$,
which is not possible since $z$ is minimal in $S \cap X$.
\end{proof}

We apply this lemma with the collection of derivations as $\XX$.
To this end
we carefully define derivations so that any derivable element
determines its down-set (see \prettyref{fig:wod}).
This led to the following definition:
\newcommand\low[2]{{#1{\downarrow}#2}}
\begin{isabelle}
\definition "derivation $X$ \eq $X$ $\subseteq$ $A$ \AND well_ordered_set $X$ (\SLE) \AND 
  (\ALL{$x$} \IN $X$. \Let $Y$ = \{$y$ \IN $X$. $y$ \SLT $x$\} \In
    (\EX{$y$}. extreme $Y$ (\SLE) $y$ \AND $x$ = $f$ $y$) \OR ($f$ \` $Y$ $\subseteq$ $Y$ \AND extreme_bound $A$ (\SLE) $Y$ $x$))"
\end{isabelle}
First, note that we demand that a derivation is well-ordered not just
well-founded.
This deviation does not make essential difference since
any derivation is proven to be connex in Grall's approach.
Second,
we demand that every $x$ in a derivation $X$ is ``derived''
from its predecessors $\low Xx \defeq \set{y\in X.\ y \SLT x}$
as either
\begin{itemize}
\item
a \emph{successor}: $\low Xx$ has a greatest element $y$ and $x = f\:y$, or
\item
a \emph{limit}:
$\low Xx$ is closed under $f$ and $x$ is a supremum of $\low Xx$.
\end{itemize}
The closure condition in the limit case
is the key trick to ensure the uniqueness of the down-set.

In the coming~\prettyref{sec:generalsetting} we provide a general condition
 which ensures the existence of a fixed point.
Afterwards we instantiate the condition to obtain
generalizations of the theorems by Bourbaki--Witt, Markowsky, Pataraia,
and Bhatta.
None of the proofs use the axiom of choice.

\subsection{General Setting}
\label{sec:generalsetting}

We first prove that derivations are downward closed,
if $f$ satisfies a variant of the inflation and reflexivity conditions on derivations:
\begin{isabelle}
\context
  \assumes derivation_infl: "\ALL{$X$ $x$ $y$}. derivation $X$ \imp $x$ \IN $X$ \imp $y$ \IN $X$ \imp
                                $x$ \SLE $y$ \imp $x$ \SLE $f$ $y$"
    \and derivation_f_refl: "\ALL{$X$ $x$}. derivation $X$ \imp $x$ \IN $X$ \imp $f$ $x$ \SLE $f$ $x$"
    \and "antisymmetric $A$ (\SLE)"
\end{isabelle}
We will show that monotone maps satisfy the first two conditions.
At this point we require antisymmetry:
incomparable successors may be derived from distinct limits,
destroying connexity. Indeed, suppose that $x$ is derivable, obtained from the successor case
$x = f\:z$ with $z$ being a greatest element of $\low Xx$, and $u$ is another 
greatest element of $\low Xx$. Then we expect $f\:u$ to be derivable, 
but it is possible that $f\;u$ and $x$ are incomparable (remember that, although $u \sim z$, we do not 
assume monotonicity at this point).
Nevertheless the condition will be relaxed to a milder condition in a later section.

The following lemma is derived from Grall's proof.
We simplify the claim so that we consider two elements from one derivation,
instead of two derivations.
\begin{isabelle}
\lemma derivation_useful:
  \assumes "derivation $X$" \and "$x$ \IN $X$" \and "$y$ \IN $X$" \and "$x$ \SLT $y$"
  \shows "$f$ $x$ \SLE $y$"
\end{isabelle}
\begin{proof}
This is done by proving the following stronger claim:
\begin{isabelle}
"($x$ \SLT $y$ \imp $f$ $x$ \SLE $y$ \AND $f$ $x$ \IN $X$) \AND ($y$ \SLT $x$ \imp $f$ $y$ \SLE $x$ \AND $f$ $y$ \IN $X$)"
\end{isabelle}
by induction on $x \in X$, and then on $y \in X$.
Remember that induction on elements of $X$ is possible because 
derivations are well-related and thus well-founded.
Let us present a proof only for the case where $x \SLT y$.
The case $y \SLT x$ is similar, while the induction hypothesis on $x$ is used instead of $y$.
The proof continues by case distinction on $y \in X$, namely, whether it is a 
successor or a limit.
\begin{itemize}
\item{Successor case:}
Suppose that there is a greatest element $u$ in $\low Xy$ and $y = f\:u$.
Since $X$ is antisymmetric and connex,
only the following three comparisons $x$ and $u$ are possible:
\begin{itemize}
\item $x \SLT u$: Using the induction hypothesis on $u \SLT y$, we know that $f\:x \SLE u$. 
Since $u \in X$, by \isa{derivation_infl}, $f\:x \SLE f\:u = y$.
\item $x = u$: we have $f\:x = y$ so $f\:x \in X$, and since $(X,\SLE)$
is well-ordered and thus reflexive, $f\:x = y \SLE y$.
\item $u \SLT x$: By the induction hypothesis on $u \SLT x$,
we have $y = f\:u \SLE x$. 
However, by assumption $x \SLT y$, and so $y \not\SLE x$, 
which is impossible.
\end{itemize}
\item{Limit case:}
Suppose that $\low Xy$ is closed under $f$ and $y$ is its supremum.
Since $x \SLT y$ we have $x \in \low Xy$, and since $\low Xy$ is closed, $f\:x \in \low Xy$.
This means
$f\:x \in X$ and $f\:x \SLT y$.\qedhere
\end{itemize}
\end{proof}

The next one is the main lemma of this section,
stating that elements from two possibly different derivations are
comparable, and moreover the lower one is in the derivation of the upper one.
The latter claim, not found in Grall's proof, is crucial in proving
that the union of all derivations is well-related.
\begin{isabelle}
\lemma derivations_cross_compare:
  \assumes "derivation $X$" \and "derivation $Y$" \and "$x$ \IN $X$" \and "$y$ \IN $Y$"
  \shows "($x$ \SLT $y$ \AND $x$ \IN $Y$) \OR $x$ = $y$ \OR ($y$ \SLT $x$ \AND $y$ \IN $X$)" 
\end{isabelle}
\begin{proof}
The proof is conducted by induction on $x \in X$ and then on $y \in Y$.
We prove $(y \SLT x \land y \in X) \vee x \SLE y$
using the induction hypothesis on $x$:
\begin{isabelle}
IHx: "($z$ \SLT $y$ \AND $z$ \IN $Y$) \OR $z$ = $y$ \OR ($y$ \SLT $z$ \AND $y$ \IN $X$)"
\end{isabelle}
for any $z \in \low Xx$.
The symmetric statement is proved similary using the induction hypothesis on $y$,
which allows us to conclude the proof.

We proceed by case distinction on $x$.
\begin{itemize}
\item
{Successor case:}
Suppose that $\low Xx$ has a greatest element $z$ and $x = f\:z$.
By \isa{IHx} we have the following three possibilities:
\begin{itemize}
\item $z \SLT y$ and $z \in Y$: by \isa{derivation_useful} in $Y$ applied to $z \SLT y$, we obtain that $x = f\:z \SLE y$.
\item $z = y$: since $z \in \low Xx$, we know $y \SLT x$ and $y \in X$.
\item $y \SLT z$ and $y \in X$: since $z \in \low Xx$, we have $z \SLT x$, 
and since $(X,\SLE)$ is a well-order, $y \SLT z \SLT x$ implies $y \SLT x$.
\end{itemize}
\item
{Limit case:}
Suppose that $\low Xx$ is closed under $f$ and $x$ is its supremum.
Let us prove our claim by the following case distinction:
\begin{itemize}
\item
Suppose that there exists $z \in \low Xx$ such that $y \SLE z$.
By \isa{IHx} we have $y \in X$. 
Furthermore, since $(X,\SLE)$ is a well-order,
$y \SLE z \SLT x$ implies $y \SLT x$.
\item
Otherwise,
for every $z \in \low Xx$, we have $y \not\SLT z$. 
So by \isa{IHx} we have $z \SLE y$ for all $z \in \low Xx$,
that is, $y$ is a bound of $\low Xx$. 
Since $x$ is least among such bounds, we conclude $x \SLE y$. \qedhere
\end{itemize}
\end{itemize}
\end{proof}

We say an element is \emph{derivable} if there exists a derivation X containing it.
\begin{isabelle}
\definition "derivable $x$ \eq \EX{$X$}. derivation $X$ \AND $x$ \IN $X$"
\end{isabelle}

\isalemma{derivations_cross_compare} ensures that
any two derivable elements are comparable,
and that
the set of derivations are downward closed, as in the assumptions of
\isalemma{closed_UN_well_founded}.
We then conclude that the set of derivable elements
\isa{\{$x$. derivable $x$\} = $\bigcup$\{$X$. derivation $X$\}}
is well-ordered.

\begin{isabelle}
\interpretation derivable: well_ordered_set "\{$x$. derivable $x$\}" "(\SLE)"
\end{isabelle}
and even that it forms a derivation.
\begin{isabelle}
\lemma derivation_derivable: "derivation \{$x$. derivable $x$\}"
\end{isabelle}

Moreover, the set of derivable elements is closed under $f$.
\begin{isabelle}
\lemma derivable_closed:
  \assumes "derivable $x$" \shows "derivable ($f\:x$)"
\end{isabelle}
\begin{proof}
Let $x \in X$ for a derivation $X$.
It is easy to see that $\low Xx \cup \set{x}$ is also a derivation,
and that $x$ is its maximum.
It is easy to check that $\low Xx \cup \set{x, f\:x}$ is also a derivation,
and hence $f\:x$ is derivable.
\end{proof}

Finally, if the set of all derivable elements has a supremum,
then it is a fixed point.
In particular,
since the set of derivable elements is well-related,
well-completeness ensures the existence of the fixed point.
\begin{isabelle}
\lemma sup_derivable_fp:
  \assumes "extreme_bound $A$ (\SLE) \{$x$. derivable $x$\} $p$"
  \shows "$f$ $p$ = $p$"
\end{isabelle}
\begin{proof}
Let $D$ denote the set of derivable elements.
Due to lemma \isa{derivable_closed}, we have $f \im D \subseteq D$.
This means $p$ is derivable via the limit case,
i.e., $p \in D$, and thus $f\:p \in D$.
Since $p$ is a bound of $D$,
we get $f\:p \SLE p$.
On the other hand, by assumption \isa{derivation_infl}
we have $p \SLE f\:p$,
concluding $f\:p = p$ by antisymmetry.
\end{proof}

\subsection{Instances}

We are left with the two assumptions \isa{derivation_infl} and
\isa{derivation_f_refl}.
One way to satisfy these assumptions is demanding them over the entire $A$
instead of all derivations.
We obtain the following generalization of the Bourbaki--Witt Theorem:
\begin{isabelle}
\theorem (\iin pseudo_ordered_set) well_complete_infl_imp_ex_fixed_point:
  \assumes "well_complete $A$ (\SLE)" \and "$f$ \` $A$ $\subseteq$ $A$" 
    \and "\ALL{$x$} \IN $A$. \ALL{$y$} \IN $A$. $x$ \SLE $y$  \imp $x$ \SLE $f$ $y$"
  \shows "\EX{$p$} \IN $A$. $f$ $p$ = $p$"
\end{isabelle}
Here we do not demand transitivity,
but a variant of inflation
\isa{"\ALL{$x$} \IN $A$. \ALL{$y$} \IN $A$. $x$ \SLE $y$ \imp $x$ \SLE $f$ $y$"}
rather than \isa{"\ALL{$x$} \IN $A$. $x$ \SLE $f$ $x$"}.
Note that the two conditions coincide in posets.
This result is also more general than Abian and Brown's version,
since well-completeness and weak chain-completeness coincide in posets.

Another way to satisfy
\isa{derivation_infl} and
\isa{derivation_f_refl} is to assume that $f$ is monotone,
obtaining the existence part of Bhatta and George's fixed point 
theorem~\cite{Bhatta05} without reflexivity.
Indeed, these assumptions then become provable.
\begin{isabelle}
\lemma mono_imp_derivation_infl:
  \assumes "monotone_on $A$ (\SLE) (\SLE) $f$"
  \shows "$\forall X\:x\:y$. derivation $X$ \imp $x \in X$ \imp $y \in X$ \imp $x \SLE y$ \imp $x \SLE f\:y$"
\end{isabelle}
\begin{proof}
Fix a derivation $X$ and $y \in X$.
We prove the claim by induction on $x$,
namely, assuming the following induction hypothesis:
\begin{isabelle}
  IH: "$z$ \SLE $y$ \imp $z$ \SLE $f$ $y$"
\end{isabelle}
for all $z \in \low Xx$, we prove that $x \SLE y$ implies $x \SLE f\:y$.
We proceed by case analysis on $x \in X$.
\begin{itemize}
\item {Successor case:}
Suppose that the greatest element $z$ in $\low Xx$ exists and $x = f\:z$.
Since $(X,\SLE)$ is well-ordered, $z,x,y \in X$ and $z \SLT x \SLE y$,
we have $z \SLE y$.
Then by monotonicity, $x = f\:z \SLE f\:y$.

\item {Limit case:}
Suppose that $\low Xx$ is closed under $f$ and $x$ is its supremum.
It is then enough to prove that $f\:y$ is a bound of $\low Xx$.
So let $z \in \low Xx$.
We have $z \SLT x \SLE y$ and as in the above case, $z \SLE y$.
By \isa{IH}, 
we get that $z \SLE f\:y$,
and we conclude by extremality of $x$.\qedhere
\end{itemize}
\end{proof}

\begin{isabelle}
\lemma mono_imp_derivation_f_refl:
  \assumes "monotone_on $A$ (\SLE) (\SLE) $f$"
  \shows "\ALL{$X$ $x$}. derivation $X$ \imp $x$ \IN $X$ \imp $f$ $x$ \SLE $f$ $x$"
\end{isabelle}
\begin{proof}
Let $X$ be a derivation and $x \in X$.
We know that $(X,\SLE)$ is well-ordered and thus reflexive.
Consequently $x \SLE x$ and we conclude $f\:x\SLE f\:x$ by 
monotonicity.
\end{proof}
So we find a fixed point if $f$ is monotone.
Moreover, in this case
we can further show that the fixed point is actually the least one.
\begin{isabelle}
\lemma mono_imp_ex_least_fp:
  \assumes "well_complete $A$ (\SLE)" \and "monotone_on $A$ (\SLE) (\SLE) $f$"
  \shows "$\exists p$. extreme $\{q \in A.\ f\:q = q\}$ (\SGE) $p$"
\end{isabelle}
\begin{proof}
Due to well-completeness we obtain the supremum $p$ of the derivable elements.
We know that $p$ is a fixed point by \isalemma{sup_derivable_fp}. It remains to prove 
that $p$ is the least one. For that, we prove that every fixed point $q$ is a bound of the 
set of derivable elements.
So let $X$ be an arbitrary derivation.
We show $x \SLE q$ for every $x \in X$ by induction on $x$.
We proceed by case distinction on $x \in X$.
\begin{itemize}
\item
{Successor case:}
Suppose that $\low Xx$ has a greatest element $z$ and $x = f\:z$.
Since $z \in \low Xx$, by the induction hypothesis we have $z \SLE q$. By 
monotonicity, we get $x = f\:z \SLE f\:q = q$.
\item
{Limit case:}
Suppose that $\low Xx$ is closed under $f$ and $x$ is its supremum.
By induction hypothesis $q$ is a bound of $\low Xx$,
and since $x$ is least among such,
we conclude $x \SLE q$.\qedhere
\end{itemize}
\end{proof}

To summarize this section, we proved the existence of fixed points for 
antisymmetric and well-complete relations. Inspired by Grall's proof 
we constructed a fixed point as the supremum of a well-related set defined 
using some derivation rules. This existence theorem has been instantiated 
to inflationary maps, leading to a generalization of the Bourbaki--Witt theorem without 
transitivity, as well as to monotone maps, leading to a generalization of the 
existence part of Bhatta--George's theorem, without reflexivity. In the latter, 
we also proved that the constructed fixed point is the least one.

\section{Completeness of (Quasi-)Fixed Points}
\label{sec:completeness}

Until now, we focused on proving the existence of (quasi-)fixed points.
However, fixed-point theorems for monotone maps are usually stronger: they state that 
the set of fixed points is complete itself.
The objective of this section is to prove this statement with as few order-theoretic 
assumptions as possible. We will first take a step towards completeness by proving 
existence of least quasi-fixed points, again limiting the usage of ordering assumptions. 

So how much can we generalize?
We first expected that the set of fixed points of inflationary maps might have a least element.
Nitpick~\cite{nitpick} found a counterexample to this hope.
\begin{exa}
Even in a complete poset,
an inflationary map may fail to have a least fixed point.
We stated \isa{(\iin partially_ordered_set)}
\begin{isabelle}
  \assumes "UNIV-complete $A$ (\SLE)" \and "$f$ \` $A$ $\subseteq$ $A$" \and "\ALL{$x$} \IN $A$. $x$ \SLE $f$ $x$"
  \shows "\EX{$p$}. extreme \{$p$ \IN $A$. $f$ $p$ = $p$\} (\SGE) $p$"
\end{isabelle} and
\isa{\nitpick}
found the following counterexample:
\begin{isabelle}
    $A$ = \{$a_1$, $a_2$, $a_3$, $a_4$\}
    $f$ = ($\lambda x$. _) ($a_1$ := $a_4$, $a_2$ := $a_2$, $a_3$ := $a_3$, $a_4$ := $a_4$)
    (\SLE) = ($\lambda x$. _)
      ($a_1$ := ($\lambda y$. _) ($a_1$ := True, $a_2$ := True, $a_3$ := True, $a_4$ := True),
       $a_2$ := ($\lambda y$. _) ($a_1$ := False, $a_2$ := True, $a_3$ := True, $a_4$ := False),
       $a_3$ := ($\lambda y$. _) ($a_1$ := False, $a_2$:= False, $a_3$ := True, $a_4$ := False),
       $a_4$ := ($\lambda y$. _) ($a_1$ := False, $a_2$ := False, $a_3$ := True, $a_4$ := True))
\end{isabelle}
Below we depict the relation $\SLE$ and the mapping $f$ below.
\newcommand\myarrow[1][->]{\mathrel{\tikz[baseline=-\the\dimexpr\fontdimen22\textfont2\relax]{\path[#1] (0,0) edge (.7,0)}}}
Here, an arrow $a_i \myarrow a_j$ means $a_i \SLE a_j$
and $a_i \myarrow[-latex, dashed, thick] a_j$ means $f\:a_i = a_j$.
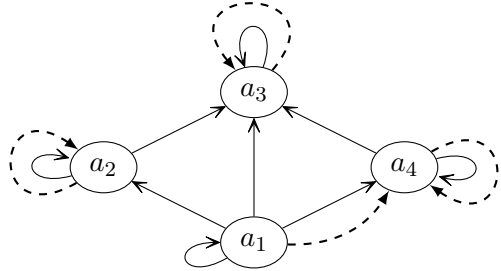
\begin{center}
\begin{tikzpicture}[scale=1]
	\node (A1) at (0,0) {$a_1$};
	\node (A2) at (-2,1) {$a_2$};
	\node (A4) at (2,1) {$a_4$};
	\node (A3) at (0,2) {$a_3$};
	\path[->]
		(A1) edge (A2)
		(A2) edge (A3)
		(A1) edge (A3)
		(A1) edge (A4)
		(A4) edge (A3)
		(A1) edge[in=-180,out=-150,looseness=8] (A1)
		(A2) edge[loop left] (A2)
		(A3) edge[in=110,out=70,looseness=8] (A3)
		(A4) edge[loop right] (A4);
	\path[-latex, dashed, thick]
		(A1) edge[bend right] (A4)
		(A2) edge[in=145,out=-150,looseness=8] (A2)
		(A3) edge[in=130,out=50,looseness=8] (A3)
		(A4) edge[in=-35,out=30,looseness=8] (A4);
\end{tikzpicture}
\end{center}
In this example, indeed $\tp{A,\SLE}$ is complete and $f$ is inflationary.
The (quasi-)fixed points are $a_2$, $a_3$, and $a_4$;
however, none of them are least: $a_2$ and $a_4$ are incomparable,
and $a_3$ is not below $a_2$ and $a_4$.
\end{exa}
So fixing our focus on monotone maps,
we try to relax ordering assumptions.
We first relaxed all ordering assumptions and asked Nitpick;
it again found a counterexample for this claim.

\begin{exa}[by Nitpick]
We stated \isa{(\iin related_set)}
\begin{isabelle}
  \assumes "UNIV-complete $A$ (\SLE)" \and "monotone_on $A$ (\SLE) (\SLE) $f$"
  \shows "\EX{$p$}. extreme \{$p$ \IN $A$. $f$ $p$ $\sim$ $p$\} (\SGE) $p$"
\end{isabelle}
Below we depict a counterexample found by \isa{\nitpick}.
Here, arrow $a_i \IFF a_j$ means $a_i \sim a_j$.
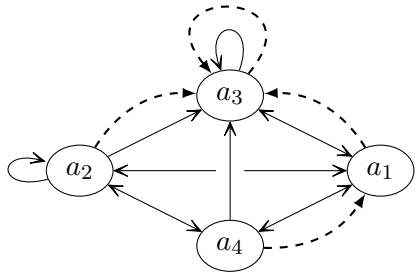
\begin{center}
\begin{tikzpicture}[scale=1]
	\node (A1) at (2,1) {$a_1$};
	\node (A3) at (0,2) {$a_3$};
	\node (A4) at (0,0) {$a_4$};
	\node (A2) at (-2,1) {$a_2$};
	\node[draw=none] (mid) at (0,1) {};
	\path[<->]
		(A2) edge (A4)
		(A1) edge (A4)
		(A1) edge (A3);
	\path[->]
		(A2) edge (A3)
		(A4) edge (A3)
		(A3) edge[in=110,out=70,looseness=8] (A3)
		(A2) edge[loop left] (A2)
		(mid) edge (A1)
		(mid) edge (A2);
	\path[-latex, dashed, thick]
		(A1) edge[bend right] (A3)
		(A2) edge[bend left] (A3)
		(A3) edge[in=130,out=50,looseness=8] (A3)
		(A4) edge[bend right] (A1);
\end{tikzpicture}
\end{center}
In this example, indeed $\tp{A,\SLE}$ is complete and $f$ is monotone.
The quasi-fixed points are $a_1$, $a_3$, and $a_4$;
however, none of them are least, because $a_1 \not\sqsubseteq a_1$, 
$a_3 \not\sqsubseteq a_4$ and $a_4 \not\sqsubseteq a_4$.
\end{exa}

After analysing the counterexample and existing proofs
for partial orders and pseudo-orders,
we found a mild requirement on $(A,\SLE)$,
that we call \emph{(semi)attractivity}:
\begin{isabelle}
\locale semiattractive = related_set +
  \assumes "$x$ $\sim$ $y$ \IMP $y$ \SLE $z$ \IMP $x$ \IN $A$ \IMP $y$ \IN $A$ \IMP $z$ \IN $A$ \IMP $x$ \SLE $z$"

\locale attractive = semiattractive +
  \assumes "semiattractive A (\SGE)"
\end{isabelle}
The intuition of this assumption is depicted in \prettyref{fig:attract}.
Attractivity is so mild that it is implied by either of antisymmetry and transitivity:
\begin{isabelle}
\sublocale transitive $\subseteq$ attractive
\end{isabelle}
\begin{isabelle}
\sublocale antisymmetric $\subseteq$ attractive
\end{isabelle}
\begin{figure}\hfil
\begin{tikzpicture}[baseline]
\draw
 (4,1) node (a) [draw] {$z$}
 (2,1) node (b) [draw] {$y$}
 (0,1) node (b') [draw] {$x$};
\path[->]
 (b) edge (a)
 (b) edge[bend right] (b')
 (b') edge[bend right] (b)
 (b'.south east) edge[dashed, bend right] (a.south west);
\end{tikzpicture}\hfil
\begin{tikzpicture}[baseline]
\draw
 (4,1) node (a) [draw] {$x$}
 (2,1) node (a') [draw] {$y$}
 (0,1) node (b) [draw] {$z$};
\path[->]
 (b) edge (a')
 (a) edge[bend left] (a')
 (a') edge[bend left] (a)
 (b) edge[dashed,out=-35,in=-135] (a);
\end{tikzpicture}\hfil
\caption{\label{fig:attract}Attractivity:
If two elements are similar, then arrows coming to one of them are also ``attracted'' to the other.
}
\end{figure}

\subsection{Least Quasi-Fixed Points for Attractive Relations}

We show now that a monotone map on a 
well-complete attractive set has a least quasi-fixed point.
For later use, we further show that the quasi-fixed point is
smaller than any \emph{strict} fixed points;
note that not all strict fixed points are quasi-fixed if we do not assume reflexivity.

Let us denote by $(\SLE^s)$ the extension of $(\SLE)$ to sets.
\begin{isabelle}
\definition "$X \SLE^s Y$ \eq $\forall x \in X.\ \forall y \in Y.\ x \SLE y$"
\end{isabelle}

\begin{isabelle}
\lemma attract_mono_imp_least_qfp:
  \assumes "attractive $A$ (\SLE)" \and "well_complete $A$ (\SLE)" \and "monotone_on $A$ (\SLE) (\SLE) $f$"
  \shows "\EX{$c$}. extreme \{$p$ \IN $A$. $f$ $p$ $\sim$ $p$ \OR $f$ $p$ = $p$\} (\SGE) $c$ \AND $f$ $c$ $\sim$ $c$"
\end{isabelle}
\begin{proof}
We reduce the claim to \isalemma{mono_imp_ex_least_fp}. 
To this end, we first take the \emph{quotient} of $A$ with 
respect to $(\sim)$ to achieve antisymmetry.
\newcommand\ecl[1]{\ensuremath{[#1]_\sim}}
We define the equivalence class $\ecl x$ for given $x$ as follows:
\begin{isabelle}
\define ecl ("[_]\sb{$\sim$}") \where "\ecl{x} \eq \{$y$ \IN $A$. $x$ $\sim$ $y$\} $\cup$ \{$x$\}" \for $x$
\end{isabelle}
Note that we explicitly include $\{x\}$ because
we do not assume reflexivity, so not necessarily $x \sim x$.
Mathematically, $\ecl{x}$ is the equivalence classe of $x$ for the equivalence
relation $(\sim) \cup (=)$. This relation is trivially symmetric and reflexive, and 
it is transitive by the attractivity of $(\SLE)$.
We collect such equivalence classes into $Q$.
Here, \isa{"\{$g$ $x$ |. $x \in A$\}"} is our notation for the set $\{g\ x \mid x \in A\}$.
\begin{isabelle}
\define $Q$ \where "$Q$ \eq \{\ecl{x} |. $x$ \IN $A$\}"
\end{isabelle}
Let us say that $x \in A$ \emph{represents} $\ecl{x}$.
The first observation is (1): any $x \in X$ represents $X \in Q$.
Indeed, if $y \in \ecl{x}$, then $y \sim x$. So for any $z \sim x$
by attractivity we have $z \sim y$, and $\ecl{x} \subseteq \ecl{y}$.
The other inclusion is symmetric.
The second observation is (2):
$\ecl{x} \SLE^s \ecl{y}$ if and only if $x \SLE y$,
which is easily proved using observation (1).

We will apply \isalemma{mono_imp_ex_least_fp} to the related set $(Q,\SLE^s)$.
To this end, we need $(Q,\SLE^s)$ to be well-complete and antisymmetric.
It is straightforward to see that $\tp{Q,\SLE^s}$ is antisymmetric
using observations (1) and (2).
To see that $\tp{Q,\SLE^s}$ is well-complete,
let $C \subseteq Q$ be well-related with respect to ($\SLE^s$).
It is easy to see that $(\bigcup C,\SLE)$ is also well-related.
Since $\tp{A,\SLE}$ is well-complete, $\bigcup C$ has a supremum $x$ in $A$. 
We show that $\ecl{x} \in Q$ is a supremum of $C$ in $\tp{Q,\SLE^s}$.
\begin{itemize}
\item $\ecl{x}$ is a bound:
Let $\ecl{y} \in C$.
Since $x$ is a bound of $\bigcup C$, we have $y \SLE x$,
and thus $\ecl{y} \SLE^s \ecl{x}$ 
by observation~(2).
\item $\ecl{x}$ is least:
Let $\ecl{z}$ be a bound of $C$ in $\tp{Q,\SLE^s}$.
We have that $z$ is a bound of $\bigcup C$. Since $x$ is least among such bounds, 
\isa{$x$ \SLE $z$}, and by observation~(2) again, $\ecl{x} \SLE^s \ecl{z}$.
\end{itemize}

Finally, we need to quotient $f$:
\begin{isabelle}
\define $\fq$ \where "$\fq$ $X$ \eq \{$y$ \IN $A$. \EX{$x$} \IN $X$. $y$ $\sim$ $f$ $x$\} $\cup$ $f$ \` $X$" \for $X$
\end{isabelle}

To apply \isalemma{mono_imp_ex_least_fp} to $(Q,\SLE^s)$ and $F$,
it remains to prove that $Q$ is 
closed under $\fq$ and that $\fq$ is monotone.
For closure, it is easy to see that
$\fq\:\ecl{x} = \ecl{f\:x}$ and hence $\fq\:\ecl{x} \in Q$.
For monotonicity,
suppose $\ecl{x} \SLE^s \ecl{y}$.
Then $x \SLE y$ and thus $f\:x \SLE f\:y$ by monotonicity of $f$.
Now we know that $f\:x \in \fq\:\ecl{x}$ and $f\:y \in \fq\:\ecl{y}$,
and by observations (1) and (2), 
$\fq\:\ecl{x} \SLE^s \fq\:\ecl{y}$.

We are now able to apply \isalemma{mono_imp_ex_least_fp} to $(Q,\SLE^s)$ and $F$,
and obtain a least fixed point $P \in Q$ of $\fq$.
We conclude by proving that any $p \in P$
is a quasi-fixed point of $f$ and that it is least among 
(quasi-)fixed points.
\begin{itemize}
\item $p$ is a quasi-fixed point: Since $p \in P$, $f\,p \in \fq\:P$.
Since $P$ is a fixed point of $\fq$, $P = \fq\:P$ and thus $p \in \fq\:P$.
Consequently, $f\:p \sim p$ or $f\:p = p$.
Since $P$ is least, we have $P \SLE^s P$, which implies that $p \SLE p$ 
and that in any case $f\:p \sim p$.
\item $p$ is least:
Let $q$ be a (quasi-)fixed point, i.e., $f\:q \sim q$ or $f\:q = q$.
Then we have $f\:q \in \ecl{q}$ and thus $\ecl{f\:q} = \ecl{q}$.
We also have $\ecl{f\:q} = \fq\:\ecl{q}$, so we conclude that $\fq\:\ecl{q} = \ecl{q}$,
that is, $\ecl{q}$ is a fixed point of $\fq$.
Since $P$ is the least fixed point of $\fq$,
we have $P \SLE^s \ecl{q}$, which implies $p \SLE q$.\qedhere
\end{itemize}
\end{proof}


\subsection{General Completeness}

Using \isa{attract_mono_imp_least_qfp},
we prove the following general completeness theorem:
Let $f$ be a monotone map over 
an attractive $\CC$-complete related set $(A,\SLE)$,
such that
$\CC$ contains all well-related subsets of $A$ and
is closed under ordered unions (\isa{extend}).
Then
the set of quasi-fixed points of $f$,
augmented with arbitrary strict fixed points,
is $\CC$-complete.

The conditions on $\CC$ are satisfied in all completeness assumptions used 
for fixed-point theorems,
as demonstrated in \prettyref{sec:ord_union}.

\begin{isabelle}
\theorem attract_mono_imp_fp_qfp_complete:
  \assumes "attractive $A$ (\SLE)" \and "$\CC$-complete $A$ (\SLE)"
    \and "\ALL{$X$} $\subseteq$ $A$. well_related_set $X$ (\SLE) \imp $X$ \IN $\CC$"
    \and extend: "$\forall X \in \CC.\ \forall Y \in \CC.\ X \SLE^s Y \imp X \cup Y \in \CC$"
    \and "monotone_on $A$ (\SLE) (\SLE) $f$" \and "$P$ $\subseteq$ \{$x$ \IN $A$. $f$ $x$ = $x$\}"
  \shows "$\CC$-complete (\{$q$ \IN $A$. $f$ $q$ $\sim$ $q$\} $\cup$ $P$) (\SLE)"
\end{isabelle}
\begin{proof}
Denote the set $\set{q \in A.\ f\:q \sim q} \cup P$ by $Q$.
  Given a subset $X$ of $Q$ in $\CC$,
  we prove that $X$
  has a supremum with respect to $\tp{Q,\SLE}$. Define the set $B$ of bounds of $X$.
\begin{isabelle}
\define $B$ \where "$B$ \eq \{$b$ \IN $A$. \ALL{$a$} \IN $X$. $a$ \SLE $b$\}"
\end{isabelle}
We first prove that $\tp{B,\SLE}$ satisfies the assumptions of \isa{attract_mono_imp_least_qfp}.
Mostly they are obvious from the corresponding assumptions on $A$ and
$B \subseteq A$,
except for:
\begin{itemize}
	\item $f \im B \subseteq B$: Let $b \in B$.
  	By the definition of $B$, for any $a \in X$ we have $a \SLE b$,
  	and with monotonicity $f\:a \SLE f\:b$.
    If $f\:a \sim a$ then by attractivity we get $a \SLE f\:b$.
    Otherwise $a \in P$, so $a = f\:a \SLE f\:b$ and thus $f\:b \in B$.
	\item $B$ is $\CC$-complete: Fix a subset $Y$ of $B$ in $\CC$. By the definition of 
	$B$, every element in $Y$ is a bound of $X$. 
	Then by \isa{extend} we know
	$X \cup Y \in \CC$.
  By the $\CC$-completeness of $A$, 
	$X \cup Y$ has a supremum $s$ in $A$.
  We prove that $s$ is a supremum of $Y$ with respect to $\tp{B,\SLE}$:
\begin{itemize}
	\item $s$ is a bound of $Y$ by construction;
	\item $s \in B$ since it is a bound of $X$ by construction;
	\item $s \SLE b$ for any bound $b$ of $Y$ in $B$,
  since $b$ is a bound of $X\cup Y$ by the definition of $B$,
  and $s$ is least among such bounds.
\end{itemize}
\end{itemize}
Consequently, by \isa{attract_mono_imp_least_qfp} applied on $(B,\SLE)$,
we find a quasi-fixed point $q \in B$ which is least among quasi- and strict fixed points in $B$.
By the definition of $Q$, $q$ is also least in $Q \cap B$.
We conclude the proof by showing that $q$ is a supremum of $X$ with respect to $\tp{Q,\SLE}$:
\begin{itemize}
	\item $q \in Q$: by construction.
	\item $q$ is a bound of $X$: by construction, $q \in B$.
	\item $q$ is least: let $p$ be another element of $Q$ which is also a 
	bound of $X$. Then $p$ is an element in $B \cap Q$, and by 
	the construction of $q$, $q \SLE p$.
\qedhere
\end{itemize}
\end{proof}

\subsection{Instances}
\label{sec:ord_union}
We instantiate the general lemma above with various classes as $\CC$,
yielding generalizations of known 
fixed-point theorems from the literature.
Note that the general lemma demands the following mild condition on $\CC$:
\begin{isabelle}
extend: "$\forall X \in \CC.\ \forall Y \in \CC.\ X \SLE^s Y \imp X \cup Y \in \CC$"
\end{isabelle}

\paragraph{Full Completeness:}
In this case we take $\CC = \isa{UNIV}$.
Then condition \isa{extend} is trivially satisfied,
and by taking $P = \emptyset$ we obtain:
\begin{isabelle}
\theorem (\iin attractive) mono_imp_qfp_complete:
  \assumes "UNIV-complete $A$ (\SLE)" \and "$f$ \` $A$ $\subseteq$ $A$" \and "monotone_on $A$ (\SLE) (\SLE) $f$"
  \shows "UNIV-complete \{$p$ \IN $A$. $f$ $p$ $\sim$ $p$\} (\SLE)"
\end{isabelle}
Moreover, when antisymmetry is assumed,
attractivity is satisfied and quasi-fixed points are fixed points.
Although fixed points may fail to be quasi-fixed without reflexivity,
by taking $P$ as the set of fixed points we obtain:
\begin{isabelle}
\theorem (\iin antisymmetric) mono_imp_fp_complete:
  \assumes "UNIV-complete $A$ (\SLE)" \and "$f$ \` $A$ $\subseteq$ $A$" \and "monotone_on $A$ (\SLE) (\SLE) $f$"
  \shows "UNIV-complete \{$p$ \IN $A$. $f$ $p$ = $p$\} (\SLE)"
\end{isabelle}
This result generalizes Stouti--Maaden and Knaster--Tarski theorems.
In contrast to the former,
we conclude the completeness of the set of fixed points,
besides relaxing reflexivity.
Compared to the Knaster--Tarski theorem,
we have relaxed transitivity and reflexivity.

\paragraph{Connex-Completeness:}
Consider now $\CC = \isa{\{$X$. connex $X$ (\SLE)\}}$:
It is also easy to see that connex sets satisfy \isa{extend},
and we obtain completeness results for attractive sets and antisymmetric sets like in the full completeness case.
We only present the statement for antisymmetry:
\begin{isabelle}
\theorem (\iin antisymmetric) mono_imp_fp_connex_complete:
  \assumes "\{$X$. connex $X$ (\SLE)\}-complete $A$ (\SLE)"
    \and "$f$ \` $A$ $\subseteq$ $A$" \and "monotone_on $A$ (\SLE) (\SLE) $f$"
  \shows "\{$X$. connex $X$ (\SLE)\}-complete \{$p$ \IN $A$. $f$ $p$ = $p$\} (\SLE)"
\end{isabelle}
This generalizes Markowsky's result~\cite{markowsky76}
by relaxing transitivity and reflexivity.
Note that for posets,
connex-completeness and chain-completeness are equivalent.

\paragraph{Pointed Directed Completeness:}

Pointed directed-complete asserts
that every directed set, possibly empty, has a supremum.
In this work, we say $(X,\SLE)$ is \emph{directed} if any pair of two elements in $X$ has a bound in $X$.
For simplicity we allow the empty set to be directed,
which is usually not the case in the literature.
\begin{isabelle}
\definition "directed $X$ (\SLE) \eq \ALL{$x$} \IN $X$. \ALL{$y$} \IN $X$. \EX{$z$} \IN $X$. $x$ \SLE $z$ \AND $y$ \SLE $z$"
\end{isabelle}
Observe that well-related sets are connex and thus directed.
Finally, to show that directed sets satisfy \isa{extend} (without reflexivity),
we need a bit of argument.
\begin{isabelle}
\lemma directed_extend:
  \assumes "directed $X$ (\SLE)" \and "directed $Y$ (\SLE)" \and "$X \SLE^s Y$"
  \shows "directed ($X$ $\cup$ $Y$) (\SLE)"
\end{isabelle}
\begin{proof}
For any $x, y \in X \cup Y$, we find $z \in X \cup Y$ such that $x \SLE z$ and $y \SLE z$.
If either $x, y \in X$ or $x, y \in Y$,
then $z$ is found immediately as $X$ and $Y$ are directed.
So suppose $x \in X$ and $y \in Y$; the other case is symmetric.
First, we obtain $z \in Y$ such that $y \SLE z$;
note that even though $y \SLE y$ may fail to hold, we can find such $z$ as an upper bound of
$\set{y,y}$.
Since $x \in X$ and $z \in Y$, by assumption we conclude $x \SLE z$.
\end{proof}
Hence now we can consider \isa{$\CC$ = \{$X$. directed $X$ (\SLE)\}}.
Again we only present the completeness result for antisymmetry:
\begin{isabelle}
\theorem (\iin antisymmetric) mono_imp_fp_directed_complete:
  \assumes "\{$X$. directed $X$ (\SLE)\}-complete $A$ (\SLE)"
    \and "$f$ \` $A$ $\subseteq$ $A$" \and "monotone_on $A$ (\SLE) (\SLE) $f$"
  \shows "\{$X$. directed $X$ (\SLE)\}-complete \{$p$ \IN $A$. $f$ $p$ = $p$\} (\SLE)"
\end{isabelle}
which generalizes Pataraia's result~\cite{pataraia97}.

\paragraph{Well Completeness:} 
Finally, we consider $\CC = \isa{\{$X$. well_related_set $X$ (\SLE)\}}$.
\begin{isabelle}
\lemma well_related_extend:
  \assumes "well_related_set $X$ (\SLE)" \and "well_related_set $Y$ (\SLE)" 
    \and "$X \SLE^s Y$"
  \shows "well_related_set ($X$ $\cup$ $Y$) (\SLE)"
\end{isabelle}
\begin{proof}
Let $Z \subseteq X\cup Y$ with $Z \neq \{\}$. 
We prove that $Z$ has a least element $z$.
We consider the following two cases: 
\begin{itemize}
	\item If $Z \cap X = \{\}$, then $Z \subseteq Y$ and $Z$ has a 
	least element $z$ since $Y$ is well-related.
	\item Otherwise, $Z \cap X \neq \{\}$ and $Z \cap X \subseteq X$. 
	Let $z$ be least in $Z \cap X$, which exists since $X$ is 
	well-related. Then $z$ is also least in $Z = (Z\cap X)\cup(Z\cap Y)$
	since $z \in X$ is below every element in $Z\cap Y \subseteq Y$ by assumption.\qedhere
\end{itemize}
\end{proof}
We then obtain the following result:
\begin{isabelle}
\theorem (\iin antisymmetric) mono_imp_fp_well_complete:
  \assumes "well_complete $A$ (\SLE)" \and "$f$ \` $A$ $\subseteq$ $A$" \and "monotone_on $A$ (\SLE) (\SLE) $f$"
  \shows "well_complete \{$p$ \IN $A$. $f$ $p$ = $p$\} (\SLE)"
\end{isabelle}
Recall that, under antisymmetry, well-ordered sets are well-related sets,
and thus weak chain\hyp completeness and well\hyp completeness coincide.
Consequently the above theorem generalizes Bhatta and George~\cite{BG11}'s theorem
by relaxing reflexivity.
Although the generalization is mild, we stress that our proof does not use ordinals (and 
is formalized in Isabelle).

All those instances witness the advantage of our approach. By proving the completeness of the set of (quasi)-fixed points \emph{as general as possible}, we obtained all such theorems we know in the literature almost for free. Each of them is a 3-to-4-line Isabelle 
proof, made even more immediate by the usage of locales.

\section{Iterative Fixed-Point Theorem}
\label{sec:kleene}

Kleene's fixed-point theorem states that,
for a pointed directed complete poset $\tp{A,\SLE}$
and a Scott-continous map $f: A \to A$,
the supremum of $\set{f^n\:\bot \mid n\in\Nat}$ exists in $A$ and is the least fixed point.
Mashburn~\cite{mashburn83} generalized the result so that
$\tp{A,\SLE}$ is an $\omega$-complete poset
and $f$ is $\omega$-continuous.

In this section we further generalize the result and show that
for any $\omega$-complete related set $\tp{A,\SLE}$
and for any bottom element $\bot \in A$,
the set $\set{f^n\:\bot \mid n\in\Nat}$ has suprema (not necessarily unique, of 
course), and these are quasi-fixed points.

\subsection{Scott Continuity, Omega-Completeness, Omega-Continuity}

We say that a related set $\tp{A,\SLE}$ is \emph{$\omega$-complete} if every $\omega$-chain---%
a chain of countably infinite cardinality---%
has a supremum.
In order to characterize $\omega$-chains in Isabelle (without going into ordinals),
we model them as the range of a relation-preserving map $c : \Nat \to A$. 
Here,\break \isa{\{$f$ $x$ | $x$ :: 'a. $P$ $x$\}}
denotes the set $\{f\ x \mid P\ x\}$, where $x$ ranges over type \isa{'a}.

\begin{isabelle}
\definition "omega_complete $A$ (\SLE) \eq 
  \{range $c$ | $c$ :: nat \To 'a. monotone ($\leq$) (\SLE) $c$\}-complete $A$ (\SLE)"
\end{isabelle}
Note here that \isa{monotone} from the Isabelle library is equivalent to
\isa{monotone_on UNIV}.
A map $f : A \to A$ is \emph{Scott-continuous} with respect to 
$(A,\SLE)$ if
for every nonempty directed subset $X \subseteq A$ with a supremum $s$,
$f\:s$ is a supremum of the image $f \im X$.
\begin{isabelle}
\definition "scott_continuous $A$ ($\SLE$) $f$ \eq $f$ \` $A$ $\subseteq$ $A$ \AND 
  ($\forall X$ $s$. $X \subseteq A$ \imp directed $X$ (\SLE) \imp $X$ $\neq$ \{\} \imp
    extreme_bound $A$ (\SLE) $X$ $s$ \imp extreme_bound $A$ (\SLE) ($f$ \` $X$) ($f$ $s$))"
\end{isabelle}
The notion of \emph{$\omega$-continuity} relaxes Scott-continuity by
considering only $\omega$-chains.

\begin{isabelle}
\definition "omega_continuous $A$ ($\SLE$) $f$ \eq $f$ \` $A$ $\subseteq$ $A$ \AND
  (\ALL{$c$} :: nat \To 'a. \ALL$s$ \IN $A$. range $c$ $\subseteq$ $A$ \imp monotone (\LE) (\SLE) $c$ \imp
    extreme_bound $A$ (\SLE) (range $c$) $s$ \imp extreme_bound $A$ (\SLE) ($f$ \` range $c$) ($f$ $s$))"
\end{isabelle}
As $\tp{\Nat,\le}$ is connex, and thus directed,
we can easily verify that Scott-continuity implies $\omega$-continuity
using the fact that the image of a monotone map over a directed set is directed.
\begin{isabelle}
\lemma scott_continous_imp_omega_continous:
  \assumes "scott_continuous $A$ ($\SLE$) $f$" \shows "omega_continuous $A$ ($\SLE$) $f$"
\end{isabelle}

For the later development we also prove that every $\omega$-continuous function is \emph{nearly} monotone, 
in the sense that it preserves relation $x \SLE y$ when
$x$ and $y$ are reflexive elements.
Note that near monotonicity coincides with monotonicity if the underlying relation is reflexive.
\begin{isabelle}
\lemma omega_continous_imp_mono_refl:
  \assumes "omega_continuous $A$ ($\SLE$) $f$" \and "$x$ \SLE $y$" \and "$x$ \SLE $x$" \and "$y$ \SLE $y$"
  \shows "$f$ $x$ \SLE $f$ $y$"
\end{isabelle}

\begin{proof}
The proof consists in observing that under the assumptions,
function \isa{$c$ :: nat \To 'a}
defined by \isa{"$c$ $i$ $\equiv$ \If $i$ = 0 \Then $x$ \Else $y$"}
is monotone.
Furthermore, $y$ is a supremum of the image of $c$,
i.e., $\set{x,y}$, so 
$\omega$-continuity ensures that $f\:y$ is a supremum of $\set{f\:x, f\:y}$, 
which in particular means that $f\:x \SLE f\:y$.
\end{proof}

\subsection{Existence of Iterative Fixed Points}

Now we prove that
if the set $\set{f^n\:\bot \mid n \in \Nat}$
has a supremum, which is implied by $\omega$-completeness,
then it is a quasi-fixed point. We prove this claim without assuming 
anything on $(A,\SLE)$ besides one bottom element.
\begin{isabelle}
\context
  \fixes $A$ \and less_eq (\infix "\SLE" 50) \and bot ("$\bot$") \and $f$
  \assumes "\ALL$x$. $\bot$ \SLE $x$" \and "omega_continuous $A$ (\SLE) $f$" 
\ibegin
\end{isabelle}

Just for convenience we abbreviate the set
$\set{f^n\:\bot \mid n \in \Nat}$ as \isa{Fn} in Isabelle.
\begin{isabelle}
\abbreviation "Fn \eq \{$f^n$ $\bot$ |. $n$ :: nat\}"
\end{isabelle}
The first observation is that \isa{Fn} is an $\omega$-chain.
In our formalization,
this means showing that \isa{Fn}
is the range of a monotone map from $\tp{\Nat,\le}$ to $\tp{A,\SLE}$.
To this end consider the mapping \isa{fn} defined by
$\isa{fn}\:i\defeq f^i\:\bot$.
Indeed, \isa{Fn = range fn} is trivial, and
monotonicity is reduced to $f^n\:\bot \SLE f^{n+k}\:\bot$
for any $n$ and $k$, which is easily proved by induction on $n$.
Hence, $\omega$-completeness yields a supremum for \isa{Fn}:
\begin{isabelle}
\lemma ex_kleene_qfp:
  \assumes "omega_complete $A$ (\SLE)" \shows "$\exists p$. extreme_bound $A$ (\SLE) Fn $p$" 
\end{isabelle}
Secondly, this supremum is a quasi-fixed point.
\begin{isabelle}
\theorem kleene_qfp:
    \assumes "extreme_bound (\SLE) Fn $p$ \shows $f\:p \sim p$"
\end{isabelle}
\begin{proof}
Since $p$ is a supremum of \isa{Fn},
the $\omega$-continuity of $f$ ensures that $f\:p$ is a supremum of $f \im \isa{Fn}$.
As $p$ is a bound of \isa{Fn},
it is also a bound of $f \im \isa{Fn} \subseteq \isa{Fn}$. Consequently, $f\:p \SLE p$.

It remains to show the other orientation $p \SLE f\:p$.
Since $p$ is least among the bounds of \isa{Fn},
it suffices to show that $f\:p$ is a bound of \isa{Fn},
that is, $f^n\:\bot \SLE f\:p$ for every $n$. We prove this by induction on $n$.
The base case is by the assumption of $\bot$.
For inductive case,
assume $f^n\:\bot \SLE p$.
Since $p$ is an extreme bound, $p \SLE p$, and by ``near'' monotonicity 
we conclude $f^{n+1}\:\bot \SLE f\:p$.
To this end we need $f^n\:\bot \SLE f^n\:\bot$ for every $n$,
which would be trivial if we had reflexivity.
Instead we prove this fact by induction on $n$, also using \isa{omega_continous_imp_mono_refl}.
\end{proof}

Now the first part of Mashburn's theorem is reproved without any order assumption:
for an $\omega$-complete set $\tp{A,\SLE}$ with a bottom element $\bot$
and $\omega$-continuous map $f : A \to A$, 
there exists a supremum for $\set{f^n\:\bot \mid n \in \Nat}$
and it is a quasi-fixed point.

\subsection{Iterative Fixed Points are Least}

Though we proved the existence of a quasi-fixed point,
Kleene's and Mashburn's fixed point theorems moreover claim that the fixed point is exactly the least one (in posets).
Hence naturally we considered proving this claim for arbitrary relations,
but again Nitpick saved us this hopeless effort.
\begin{exa}[by Nitpick]
Our conjecture now \isa{\assumes "extreme_bound (\SLE) Fn $q$"}
and \isa{\shows "extreme (\SGE) \{$s$. $f$ $s$ $\sim$ $s$\} $q$"}.
Following we depict a counterexample found by \isa{\nitpick}:
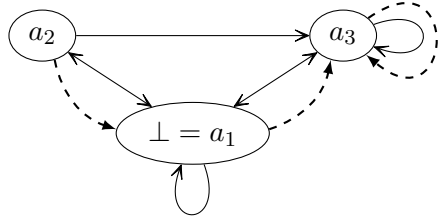
\begin{center}
\begin{tikzpicture}[scale=1]
	\node (A1) at (2,0) {$\bot = a_1$};
	\node (A3) at (4,1.3) {$a_3$};
	\node (A2) at (0,1.3) {$a_2$};
	\path[->]
		(A2) edge (A3)
		(A3) edge[in=-20,out=20,looseness=8] (A3)
		(A1) edge[in=-110,out=-70,looseness=8] (A1);
	\draw[<->]
		(A2) edge (A1)
		(A1) edge (A3);
	\draw[-latex, dashed, thick]
		(A1) edge[bend right] (A3)
		(A2) edge[bend right] (A1)
		(A3) edge[in=-40,out=35, looseness=8] (A3);
\end{tikzpicture}
\end{center}
In this example,
indeed $a_1$ is a bottom element, $\SLE$ is ($\omega$-)complete,
and $f$ is $\omega$-continuous.
The set of quasi-fixed points is $\{a_1, a_2, a_3\}$,
and $a_3$ is a supremum of $\set{f^n\:\bot \mid n \in \Nat} = \{a_1, a_3\}$. However, $a_3$ is not a least quasi-fixed point because $a_3 \not\sqsubseteq a_2$.
\end{exa}

Now again, attractivity turns out to be the key.
\begin{isabelle}
\theorem(\iin attractive) kleene_qfp_is_dual_extreme:
  \assumes "omega_complete $A$ (\SLE)" \and "omega_continuous $A$ (\SLE) $f$" 
    \and "$\bot \in A$" \and "$\forall x \in A.\ \bot \SLE x$"
  \shows "extreme_bound $A$ (\SLE) \{$f^n\:\bot$ |. $n$ :: nat\} = extreme \{$s \in A$. $f\:s \sim s$\} (\SGE)"
\end{isabelle}

\begin{proof}
Let $q$ be a supremum of \isa{Fn}. By \isa{kleene_qfp}, 
we already know that this is a quasi-fixed point.
So to prove that $q$ is a least quasi-fixed point,
it is enough to show that
any other quasi-fixed point $s$ is a bound of $\isa{Fn} = \set{f^n\:\bot \mid n\in\Nat}$.
This is done by induction on $n$.
The base case $\bot \SLE s$ is trivial by assumption.
For the inductive case, assuming $f^n\:\bot \SLE s$ we get $f^{n+1}\:\bot \SLE f\:s$
by the same argument as in the previous proof.
Since $f\:s \sim s$,
attractivity concludes $f^{n+1}\:\bot \SLE s$.

Conversely, consider a least quasi-fixed point $s$.
We show that $s$ is a supremum of \isa{Fn}.
Since $s$ is a quasi-fixed point,
and as we have just proved above,
$s$ is a bound of \isa{Fn}.
It remains to prove that $s$ is least in bounds of \isa{Fn}.

By \isa{ex_kleene_qfp}, \isa{Fn} has a supremum $k$,
and $k$ is a quasi-fixed point.
As $s$ is a least quasi-fixed point, we have $s \SLE k$.
On the other hand,
as $s$ is a bound of \isa{Fn} and $k$ is a least of such, we see $k \SLE s$.
Consequently, $s \sim k$.

Now let $x$ be a bound of \isa{Fn}.
We know $k \SLE x$, and with $s \sim k$, 
we conclude $s \SLE x$ due to attractivity.
\end{proof}

\section{Conclusion}
In this paper, we developed an Isabelle/HOL formalization for order-theoretic fixed-point theorems.
We adopt an as-general-as-possible approach,
so that many results previously known only for partial orders or pseudo-orders are generalized 
to attractive or antisymmetric relations.
In particular, the proof of existence of a fixed point using a proof-tree-like method, as well as 
the general method to prove the completeness of the set of (quasi-)fixed points, allowed us 
to recover and generalize many known fixed-point theorems from the literature. 
These achievements become reachable to us largely due to the great assistance by
the smart Isabelle 2020 environment.

For future work,
it is tempting to further formalize and hopefully generalize other results about completeness 
and fixed points. For example, we are considering some results proved in 
\cite{markowsky76}, such as the equivalence of 
chain and pointed directed completeness, and the converse of Markowsky's fixed-point theorem,
both requiring some form of axiom of choice.
We also plan to extend the library with convergence arguments and to apply this general 
theory of fixed points to a domain like term rewriting, which was actually our original motivations 
for formalizing these order-theoretic concepts.

\section*{Acknowledgment}
  \noindent This work is partly supported by ERATO HASUO
Metamathematics for Systems Design Project (No.~JPMJER1603), JST and
Grant-in-aid No.~19K20215, JSPS.


\bibliographystyle{alphaurl}
\bibliography{binrellmcs}

\end{document}